\documentclass[letterpaper, 10 pt, conference]{ieeeconf} 
\IEEEoverridecommandlockouts                              
\usepackage{amsmath}
\usepackage{mathtools}
\usepackage{amsfonts}
\usepackage{amssymb}
\usepackage{algorithm} 
\usepackage{algpseudocode}
\usepackage{color}
\let\labelindent\relax
\usepackage{enumitem}
\usepackage{dblfloatfix}
\usepackage[keeplastbox]{flushend}
\usepackage[normalem]{ulem}

\newtheorem{definition}{Definition}

\newtheorem{remark}{\bf Remark}

\newtheorem{theorem}{Theorem}

\newcommand{\xddots}{%
  \raise 4pt \hbox {.}
  \mkern 10mu
  \raise 1pt \hbox {.}
  \mkern 10mu
  \raise -2pt \hbox {.}
}

\newcommand{\xddotswide}{%
  \raise 5pt \hbox {.}
  \mkern 10mu
  \raise 2pt \hbox {.}
  \mkern 10mu
  \raise -1pt \hbox {.}
}

\title{\LARGE \bf Mixed Voltage Angle and Frequency Droop Control for Transient Stability of Interconnected Microgrids with Loss of PMU Measurements}
\author{S. Sivaranjani\textsuperscript{*}, Etika Agarwal\textsuperscript{*}, Le Xie, Vijay Gupta and Panos Antsaklis
\thanks{\textsuperscript{*}These authors contributed equally to this work.\newline S. Sivaranjani and Le Xie are with the Department of Electrical and Computer Engineering, Texas A\&M University, College Station, TX; \{sivaranjani,le.xie\}@tamu.edu. Etika Agarwal is with GE Global Research, India; \{etika.agarwal09@gmail.com\}. S. Sivaranjani and Etika Agarwal were with the Department of Electrical Engineering, University of Notre Dame, Notre Dame, IN when this work was carried out. Vijay Gupta and Panos Antsaklis are with the Department of Electrical Engineering, University of Notre Dame, Notre Dame, IN; \{vgupta2,pantsakl\}@nd.edu.  S. Sivaranjani and Vijay Gupta were partially supported by NSF grant CNS-1739295 and PTX-Sandia National Lab grant PO 2079716. S. Sivaranjani was also partially supported in this work by the Sclumberger Foundation Faculty for the Future fellowship. Le Xie was supported in part by NSF grants OAC-1934675, ECCS-1839616, ECCS-1611301, and CCF-1934904.  Etika Agarwal and Panos Antsaklis were partially supported by the Army Research Office Grant No. ARL W911NF-17-1-0072.
}
}

\begin{document}
\maketitle
\thispagestyle{empty}
\pagestyle{empty}

\begin{abstract}
We consider the problem of guaranteeing transient stability of a network of interconnected angle droop controlled microgrids, where voltage phase angle measurements from phasor measurement units (PMUs) may be lost, leading to poor performance and instability. In this paper, we propose a novel mixed voltage angle and frequency droop control (MAFD) framework to improve the reliability of such angle droop controlled microgrid interconnections. In this framework, when the phase angle measurement is lost at a microgrid, conventional frequency droop control is temporarily used for primary control in place of angle droop control. We model the network of interconnected microgrids with the MAFD architecture as a nonlinear switched system. We then propose a dissipativity-based distributed secondary control design to guarantee transient stability of this network under arbitrary switching between angle droop and frequency droop controllers. We demonstrate the performance of this control framework by simulation on a test 123-feeder distribution network.
\end{abstract}

\section{Introduction}
The large-scale integration of renewable and distributed energy resources (DERs) into the power grid has led to an architecture where several DERs, storage units and local loads are aggregated into clusters known as microgrids. These microgrids exchange power with each other in order to compensate for transient conditions like intermittency in renewable generation and mismatches between local generation and loads. 
The problem of designing controllers to guarantee stability of individual microgrids under transient conditions has been extensively studied (see \cite{olivares2014trends} and the references therein). 
Recently, control designs for transient stabilization of interconnected microgrids have gained attention \cite{majumder2010power}-\nocite{simpson2013stability}\nocite{zhang2016interactive}\nocite{zhang2016transient}\nocite{zamora2016multi}\nocite{li2017networked}\nocite{schiffer2016voltage}\cite{song2017distributed}.

Typically, interconnected microgrids are controlled in a hierarchical manner with three levels of control  \cite{guerrero2011hierarchical} - (i) primary control using decentralized droop settings to regulate the real and reactive power outputs of individual microgrids, (ii) secondary control to stabilize the network of droop controlled microgrids by eliminating the frequency and voltage deviations, and (iii) tertiary control to generate power references to coordinate power exchange in the network. In traditional power grids with synchronous generators, the primary control level comprises of frequency and voltage droop controllers to regulate the real and reactive power outputs respectively. However, in microgrids with high renewable energy penetration, where synchronous generators are replaced by power electronic interfaced asynchronous generators, fast-acting voltage angle droop controllers have emerged as an alternative to frequency droop controllers due to their superior transient performance and stability \cite{majumder2009angle}.


The reliability of angle droop controllers is contingent upon the availability of voltage angle measurements from phasor measurement units (PMUs). This requires an accurate Global Positioning System (GPS) signal for synchronization of the angle reference across the system \cite{sexauer2013phasor}. However, recent studies have demonstrated that PMU GPS signals are frequently lost due to factors like weather events and communication failure, leading to loss of angle measurements \cite{yao2016impact}\cite{yao2017gps}. In fact, such PMU measurement losses have been observed to occur as often as 6-10 times a day, with each loss event ranging from an average of 6-8 seconds to over 25 seconds \cite{huang2016data}. 
 Such a loss of angle measurements can  result in poor transient performance and even instability in angle droop controlled microgrids. In this context, \textit{the aim of our work is to develop a control strategy to ensure transient stability of interconnected angle droop controlled microgrids under intermittent PMU angle measurement losses}.

We present a \textit{novel mixed voltage angle and frequency droop control framework} where traditional frequency droop controllers are used to indirectly regulate real power in those microgrids where angle measurements are lost. This framework is implemented in two stages. In the first stage, the primary angle droop controller is temporarily switched to a traditional frequency droop controller in microgrids where angle measurements are lost. We model the network of interconnected microgrids in this framework as a nonlinear switched system where each microgrid switches between angle droop or frequency droop controllers, depending on the availability of PMU angle measurements. In the second stage, we ensure transient stability of the network by designing a dissipativity-based \cite{hill1980dissipative} 
 \textit{distributed secondary controller} that uses a combination of angle and frequency measurements from neighboring microgrids depending on availability. We propose a dissipativity-based controller since such designs \cite{agarwal2017feedback}-\nocite{zhang2015online}\nocite{sivaranjaniconic}\nocite{caliskan2014compositional}\nocite{agarwal2019distributed}\cite{schiffer2014conditions} have been shown to improve stability margins and transient performance in both traditional and microgrid-based network architectures.
The contributions of this paper are: 
\begin{itemize}[leftmargin=0.3cm]
    \item First, we provide a novel control framework to guarantee transient stability of a network of angle droop controlled microgrids under intermittent PMU angle measurement losses. We achieve this via a mixed voltage angle and frequency droop control (MAFD), wherein the loss of voltage angle measurements is compensated for by the use of frequency measurements. Typically, measurement losses are handled in a networked control systems framework, where the controller uses the last available measurement in the event of a measurement loss, and the maximum allowable duration of the loss is assumed to be bounded to guarantee stability \cite{sivaranjani2013networkedisgt}\cite{sivaranjani2013networked}. However, in distribution-level microgrids, the duration of measurement loss may exceed the maximum allowable duration to guarantee stability. In contrast, the MAFD framework exploits the physics of the system (that is, the fact that frequency measurements, unlike angle measurements, do not require GPS synchronization) to indirectly control the voltage angle using the frequency when angle measurements are lost. 
    \item Second, we derive a switched nonlinear dynamical model for microgrids in the MAFD setting. This model differs from traditional continuous dynamical models of interconnected angle droop controlled microgrids \cite{zhang2016transient}, where all microgrids operate in the angle droop control mode even when angle measurements are lost. In contrast, for interconnected microgrids with MAFD, the dynamics of each microgrid switches between angle droop control and frequency droop control modes depending on the availability of PMU measurements, and all microgrids may or may not operate in the same mode at the same time.
    \item Finally, we design a dissipativity-based distributed secondary controller that guarantees the stability of the interconnected microgrids even when angle measurements are lost and the primary controller is switching between angle droop and frequency droop control modes. The control design is provided in the form of linear matrix inequalities (LMIs) that guarantee stability of the nonlinear switched MAFD model based on its linear approximation. This result extends the passivity-based state-feedback control design for nonlinear discrete-time switched systems in \cite{agarwal2017feedback} to a more general continuous-time output-feedback dissipativity framework. 
\end{itemize}
\vspace{0.1em}
\textit{Notation:} $\mathbb{R}$, $\mathbb{R}_{+}$ and $\mathbb{R}^{n}$ denote the sets of real numbers, positive real numbers including zero, and $n$-dimensional real vectors respectively. Given matrix $A\in\mathbb{R}^{m\times n}$, $A_{ij}$ and $A'$ represent its $(i,j)$-th element and its transpose respectively. The identity matrix is denoted by $I$, with dimensions clear from the context. A symmetric positive definite (semi-definite) matrix $P\in\mathbb{R}^{n \times n}$ is represented as $P{>}0$ ($P{\geq}0$). For a set $B$, $|B|$ denotes its cardinality. Given two sets $A$ and $B$, $A\backslash B$ represents the set of all elements of $A$ that are not in $B$. For a complex number $z{=}a+b\sqrt{-1}, a,b \in \mathbb{R}$, $|z|{=}\sqrt{a^2+b^2}$ and $\angle z{=}\arctan({b}/{a})$ represent its magnitude and phase angle respectively. For a set $\Sigma{=}\{a,b\}$, $a,b\in\mathbb{R}$ and $n\in\mathbb{R}_+\backslash\{0\}$, $\Sigma^{n}$ represents the set of all possible $v \in \mathbb{R}^n$, whose $i$-th component $[v]_i\in\Sigma$, $i\in1,..,n$.

\section{Preliminaries}
Consider a network of $N$ microgrids where each microgrid is connected to the network through a {voltage-source inverter (VSI)-based} power-electronic interface. The bus at which each microgrid is interfaced to the network is known as its point of common coupling (PCC). The topology of this network is given by a weighted undirected graph $\mathcal{G}(\mathcal{V},\mathcal{E})$, where nodes $\mathcal{V}$ represent PCCs, edges $\mathcal{E}$ represent power transmission lines connecting these PCCs, and $|\mathcal{V}| = N$. 
The {adjacency matrix} for this network is denoted by $T\in \mathbb{R}^{N\times N}$, where $T=[T_{jk}]$, $j,k \in \{1,2,\ldots,N\}$, with $T_{jk}=1$ if there exists an edge connecting node $j$ to $k$ and $T_{jk}=0$ otherwise. We assume $T_{jj}=1$, that is, every node has a self-loop. To every edge between nodes $j$ and $k$, we assign a weight $Y_{jk}$, representing the complex admittance of the line between $j$ and $k$.  Similarly, the self admittance at node $j \in \{1,2,\ldots,N\}$ is denoted by $Y_{jj}$. The \textit{set of neighbors} of microgrid $j$ is denoted by $\mathcal{N}_j=\{k \in \{1,2,\ldots,N\}: T_{jk}=1\}$. Using the standard AC power flow model, the net real and reactive power injections, $P_{inj}^j(t)$ and $Q_{inj}^j(t)$, at the $j$-th bus at time $t$ are given by
{\small\begin{equation}\label{powerflow}
\begin{aligned}
P_{inj}^j(t)&= \sum \limits_{k \in \mathcal{N}_j} V_j(t) V_k(t) |Y_{jk}| \sin(\delta_{jk}(t)+\pi /2-\angle{Y_{jk}}) \\
Q_{inj}^j(t)&= \sum \limits_{k \in \mathcal{N}_j} V_j(t) V_k(t) |Y_{jk}| \sin(\delta_{jk}(t)-\angle{Y_{jk}}),
\end{aligned}
\end{equation}}
where $V_j(t)$ and $\delta_j(t)$ are the voltage magnitude and phase angle at the $j$-th bus respectively, and $\delta_{jk}(t)=\delta_j(t)-\delta_k(t)$.

\subsection{Conventional Control}\label{sec:conventional}
Stability of interconnected microgrids is typically guaranteed by hierarchical control \cite{guerrero2011hierarchical} comprised of (i) a primary control layer to ensure proper load sharing between microgrids, and (ii) a secondary control layer to ensure system stability by eliminating frequency and voltage deviations. 
	
	\textbf{Frequency Droop Control:} Traditionally, the real and reactive power injections are regulated to desired set points (to compensate for the generation-load mismatch) by frequency droop and voltage droop controllers respectively, termed as primary controllers. For every $i \in \mathcal{V}$, the error dynamics of the microgrid (with frequency and voltage droop control) connected to bus $i$ are given by
	\begin{align}
	\Delta \dot \delta_i(t) &= \Delta \omega_i(t) \label{angle}\\
	J_{\omega_i}\Delta \dot \omega_i(t) &= -D_{\omega_i} \Delta \omega_i(t) + \Delta P_{ext}^i(t) -\Delta P_{inj}^i(t) \label{frequency}\\
	J_{V_i} \Delta \dot V_i(t) &= -D_{V_i} \Delta V_i(t) +\Delta Q_{ext}^i(t) -\Delta Q_{inj}^i(t), \label{voltage}
	\end{align}
	where $\Delta \delta_i(t)=\delta_i(t)-\delta_i^{ref}$, $\Delta \omega_i(t)=\omega_i(t)-\omega_i^{ref}$, $\Delta V_i(t)=V_i(t)-V_i^{ref}$, $\Delta P_{inj}^i(t)=P_{inj}^i(t)-P_{inj}^{i,ref}$ and $\Delta Q_{inj}^i(t)=Q_{inj}^i-Q_{inj}^{i,ref}$ are the deviations of the angle, frequency, voltage and real and reactive power injections at the $i$-th bus from their reference values $\delta_i^{ref}$, $\omega_i^{ref}$, $V_i^{ref}$, $P_{inj}^{i,ref}$ and $Q_{inj}^{i,ref}$ respectively, $\Delta P_{ext}^i(t)$ and $\Delta Q_{ext}^i(t)$ represent the mismatch between the net generation and load at the $i$-th microgrid, and  $J_{\omega_i}$, $D_{\omega_i}$, $J_{V_i}$ and $D_{V_i}$ represent the equivalent inertia and damping coefficients corresponding to the frequency and voltage control loops. Note that \eqref{powerflow} holds with $P_{inj}^i(t)=P_{inj}^{i,ref}$, $Q_{inj}^i(t)=Q_{inj}^{i,ref}$, $V_j(t)=V_j^{ref}$, $V_k(t)=V_k^{ref}$ and $\delta_{jk}(t)=\delta_{j}^{ref}-\delta_k^{ref}$. 
	
	\textbf{Angle Droop Control:} The frequency droop control scheme regulates the real power injection in \eqref{powerflow} by indirect control of the angle via \eqref{angle}-\eqref{frequency}. However, frequency droop control has been demonstrated to suffer from issues like slow transient response and frequency drifts \cite{majumder2009angle}. Therefore, in VSI-interfaced microgrids, direct control of the voltage angle by fast-acting power electronics has emerged as an attractive alternative to classical frequency droop control \cite{majumder2010power}\cite{zhang2016transient}. In this paper, we employ the angle droop control scheme in \cite{zhang2015online}, where the error dynamics of the microgrid connected to every $i \in\mathcal{V}$ at time $t$ are described by  
	\begin{align}
	J_{\delta_i}\Delta \dot \delta_i(t) &= -D_{\delta_i} \Delta \delta_i(t) + \Delta P_{ext}^i(t) -\Delta P_{inj}^i(t)\label{angle1}\\
	J_{V_i} \Delta \dot V_i(t) &= -D_{V_i} \Delta V_i(t) +\Delta Q_{ext}^i(t) -\Delta Q_{inj}^i(t), \label{voltage-angledroop}
	\end{align}
	where $J_{\delta_i}$ and $D_{\delta_i}$ are the equivalent inertia and damping coefficients of the $i$-th microgrid with angle droop control.

	\textbf{Secondary Control:} 
The voltage magnitude and angle (or frequency) deviations caused by generation-load mismatches in a microgrid are eliminated by means of a secondary controller that uses measurements of voltage magnitudes and phase angles (or frequency) and regulates them to the desired reference \cite{guerrero2011hierarchical}. In the case of primary angle droop control, the secondary controller relies on real-time measurements of voltage magnitude and phase angle by {phasor measurement units (PMUs)} at the PCC of each microgrid. If the primary control is based on frequency droop, secondary controller then relies on real-time measurements of frequency instead of the voltage phase angle. 

	\textbf{Motivation:} 
In microgrids with primary angle droop control, the secondary controller relies on real-time measurement of voltage angles by PMUs at the PCCs, which in turn requires a GPS signal to provide an accurate reference for synchronization. Thus, angle measurements may be frequently lost due to weather and atmospheric events affecting the GPS signal \cite{yao2016impact}\cite{huang2016data}. Due to the high sensitivity and fast-acting dynamics of the angle droop control loop, a loss of angle measurement may lead to poor transient performance and even instability in the network. Motivated by this problem, the \textit{aim of this paper is to design controllers to ensure transient stability of interconnected angle-droop controlled microgrids when PMU angle measurements are lost}. 

\section{Mixed Angle and Frequency Droop Control (MAFD) Model and Transient Stability Problem} \label{sec:mafd-model}
\begin{figure}[b]
    \centering
    \includegraphics[scale=0.28,trim=0.5cm 1cm 0cm 1cm]{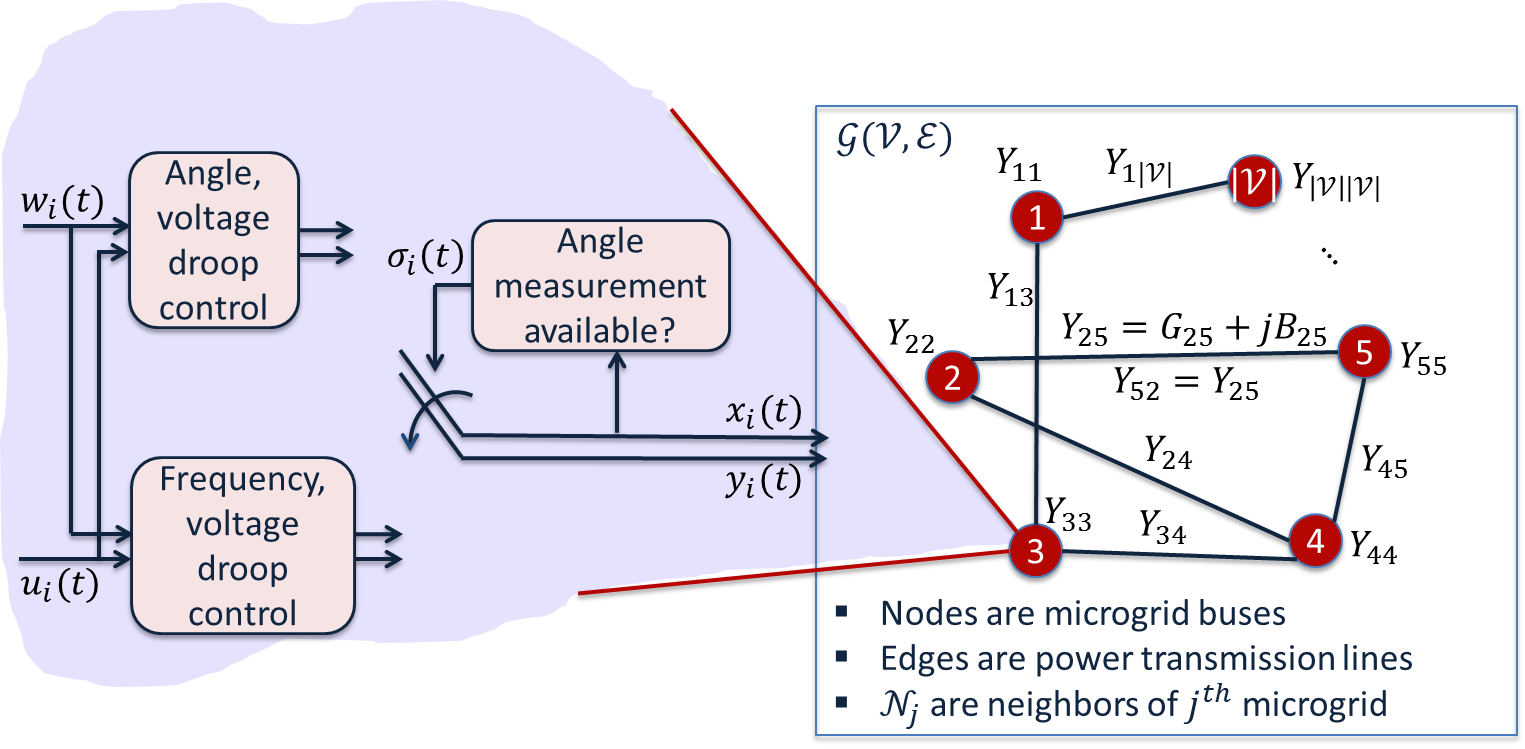}
    \caption{Mixed voltage angle and frequency droop control (MAFD) framework for interconnected microgrids}
    \label{fig:mafd-schematic}
\end{figure}

\vspace{-0.1em}
In this section, we introduce a new mixed angle and frequency droop control (MAFD) framework for primary control, to regulate the real and reactive powers of a microgrid to the desired set-points when the microgrid is subject to intermittent loss of angle measurements. The proposed scheme temporarily uses traditional frequency droop controllers for primary control in lieu of angle droop control at the microgrids where angle measurements are lost, until those measurements are restored.  Therefore, at any given time, \textit{some microgrids may operate with angle droop control while others operate with frequency droop control} depending on the availability of the angle measurements from PMU. We now formulate this MAFD framework, summarized in Fig. \ref{fig:mafd-schematic}, as a switched system model.

For every $i \in\mathcal{V}$, we define a switching signal $\sigma_i(t):\mathbb{R}_{+} \to \Sigma,$ where $\Sigma=\{1,2\}$ as the set of admissible switching values. At every time $t$, the value of $\sigma_i(t)$ represents one of the two modes in which the $i$-th microgrid in the MAFD framework is operating - (i) angle droop control mode $(\sigma_i(t) =1)$, when real-time angle measurements are available from the PMU at that microgrid, or (ii) frequency droop control mode $(\sigma_i(t) = 2)$, when PMU voltage angle measurements are lost or corrupted at that microgrid due to GPS signal loss or sensor malfunction (Fig. \ref{fig:mafd-schematic}). Since the event of loss of availability of angle measurements is not known in advance, the switching signal $\sigma_i(t)$ is also not known a priori. However, we assume that its instantaneous value is available in real-time. The dynamics of the $i\text{-th}$ microgrid in each of the modes is described as:
\begin{equation}\label{switchedsystem_state}
\vspace{-0.1em}
\begin{aligned}
\dot{x}_i(t) = f_{\sigma_i(t)}^i(x_i(t),u_i(t),w_i(t)), \quad
u_i(t) = h^i(x_i(t)),
\vspace{-0.1em}
\end{aligned}
\end{equation}
where $x_i(t)=[\Delta \delta_i(t) \; \Delta\omega_i(t) \; \Delta V_i(t)]'$, $u_i(t)=[\Delta P_{inj}^i(t) \quad \Delta Q_{inj}^i(t)]'$, $w_i(t)=[\Delta P_{ext}^i(t) \quad \Delta Q_{ext}^i(t)]'$. The output $h^i(x_i(t))$ given by \eqref{powerflow} is independent of $\sigma_i(t)$, since the power flow equations do not change with loss of angle measurements. We now describe the system dynamics $f_{\sigma_i(t)}^i(x_i(t),u_i(t),w_i(t))$ for two modes of operation of \eqref{switchedsystem_state}.

\noindent\textbf{Angle Droop Control Mode, $\mathbf{\sigma_i(t)=1}$:} This is the normal mode of operation when PMU angle measurements at the $i^{th}$ microgrid are available. The switching signal is given by $\sigma_i(t)=1$, and the dynamics $f^i_1(x_i(t),u_i(t),w_i(t))$ are described by \eqref{angle1} and \eqref{voltage-angledroop}, along with the frequency error 
\begin{align}
\Delta \dot \omega_i(t)&=-\frac{D_{\delta_i}}{J_{\delta_i}}\left[-\frac{D_{\delta_i}}{J_{\delta_i}}\Delta \delta_i(t)+\frac{1}{J_{\delta_i}}\Delta P_{ext}^i(t)\right. \label{continuity} \\ &\left. -\frac{1}{J_{\delta_i}}\Delta P_{inj}^i(t)\right]-\frac{1}{J_{\delta_i}}\Delta \dot P_{inj}^i(t), \nonumber
\end{align}
where $\Delta \dot P_{inj}^i(t)$ is the derivative of $\Delta P_{inj}^i(t)$ with respect to time $t$, computed from \eqref{powerflow}. 

\vspace{2mm}
\noindent\textbf{Frequency Droop Control Mode, $\mathbf{\sigma_i(t)=2}$:} When $\sigma_i(t)=2$, that is, in the absence of angle measurements, the frequency droop mode is employed and the system dynamics $f^i_2(x_i(t),u_i(t),w_i(t))$ are described by \eqref{angle}-\eqref{voltage}.


\begin{remark}
Typical angle droop control models ignore the dynamics of the frequency error since experimental studies have demonstrated that the variations in frequency are negligible with angle droop designs \cite{majumder2010power}\cite{majumder2009angle}. However, in contrast to these traditional models, we propagate the dynamics of the frequency error $\Delta \omega_i(t)$ through \eqref{continuity} even in the angle droop control mode in the MAFD framework for the following reason. 
To ensure continuity of the state $x_i(t)$, \eqref{angle} must be satisfied at every switching instant. This is automatically ensured for switches from the frequency droop mode to the angle droop mode. For switches from the angle droop mode to the frequency droop mode, \eqref{continuity} is sufficient to ensure continuity of the state at the switching instant, since it enforces $\Delta \dot \omega_i(t)=\Delta \ddot{\delta_i}(t)$ for all time~$t$. 
\end{remark}
\begin{figure*}[b]
\vspace{-1.1em}
\scriptsize
\hrulefill
\vspace{-0.6em}
	\newcounter{MYtempeqncnt}
    \setcounter{MYtempeqncnt}{\value{equation}}
    \setcounter{equation}{10}
{\small\begin{subequations}\label{control_lmi}
\begin{equation}
\label{pass_lmi1}
\begin{bmatrix}
    -P(A_j+B_j^{(1)}H)-(A_j+B_j^{(1)}H)'P-B_j^{(1)}U_jC_j-C_j'U_j'B_j^{(1)'} & -PB_j^{(2)}-B_j^{(1)}U_jD_j+C_j'S_j & -C_j'Q_{j-}^{1/2} \\
    -B_j^{(2)'}P-D_j'U_j'B_j^{(1)'}+S_j'C_j & D_j'S+S_j'D_j+R_j & -D_j'Q_{j-}^{1/2} \\
    -Q_{j-}^{1/2}C_j & -Q_{j-}^{1/2}D_j & I 
    \end{bmatrix}>0
\end{equation}
\begin{equation}\label{pass_lmi2}
PB_j^{(1)} = B_j^{(1)}V_j, \quad Q_{j-}^{1/2}Q_{j-}^{1/2}=-Q_j, \quad
				V_j \in \mathcal{S}_v, \quad U_j \in \mathcal{S}_H,
				\vspace{-0.9em}
\end{equation}
\end{subequations}}
\setcounter{equation}{\value{MYtempeqncnt}}
\end{figure*}

We define the augmented state, input and disturbance vectors for the network of interconnected microgrids as $x(t){=}[x_1'(t), x_2'(t), \ldots, x_{N}'(t)]'$, $u(t){=}[u_1'(t), u_2'(t), \ldots, u_{N}'(t)]'$ and $w(t)=[w_1'(t), w_2'(t), \ldots, w_{N}'(t)]'$ respectively. We also define an output vector $y_i(t)$ for every $i \in \mathcal{V}$ as 
$y_i(t)=g_{\sigma_i(t)}^i(x_i(t),w_i(t))$, 
where $g_{\sigma_i(t)}^i(t)=[\Delta \dot \delta_i(t) \; \Delta V_i(t)]'$ when $\sigma_i(t)=1$ and $g_{\sigma_i(t)}^i(t)=[\Delta \dot \omega_i(t) \; \Delta V_i(t)]'$ when $\sigma_i(t)=2$. The augmented output vector is $y(t){=}[y_1'(t), y_2'(t), \ldots, y_{N}'(t)]'$. Note that at any given time, outputs of each microgrid are the quantities whose measurements are available at that time.

With the augmented switching vector $\sigma(t)=[\sigma_1(t),\cdots,\sigma_{N}(t)]' \in \Sigma^N$, where every element can take values of 1 or 2, indicating the availability or loss of PMU angle measurement at that microgrid, we can write the dynamics of the interconnected microgrids in the MAFD framework as the nonlinear switched system
\begin{equation}\label{nonlinear_switched}
	\begin{aligned}
	\dot{x}(t) &= f_{\sigma(t)}(x(t),u(t),w(t))\\
	y(t) &= g_{\sigma(t)}(x(t),w(t))\\
    u(t) & = h(x(t)),
	\end{aligned}
\end{equation}
where $f_{\sigma(t)}=[
f^1_{\sigma_1(t)}, \ldots,
f^{N}_{\sigma_{N}(t)}
]'$, $g_{\sigma(t)}=[g^1_{\sigma_1(t)},\ldots,
g^{N}_{\sigma_{N}(t)}]'$, $h=[h^{1},\ldots,
h^{N}]'$.
Note that the origin $x(t)=0$ is an equilibrium of \eqref{nonlinear_switched}. 
We linearize each mode $j \in \Sigma^N$ of \eqref{nonlinear_switched} around the origin to obtain the linear switched system model
\begin{subequations}\label{linear switched system}
\begin{equation}
	\begin{aligned}
	\dot{{x}}(t) & = A_{\sigma(t)}{x}(t) + B_{\sigma(t)}^{(1)}{u}(t) + B_{\sigma(t)}^{(2)}{w}(t)\\
	{y}(t)&  = C_{\sigma(t)}{x}(t) + D_{\sigma(t)}{w}(t)\\
    u(t) & =Hx(t),
	\end{aligned}
	\end{equation}
	\vspace{-3pt}
   \begin{align}\label{linearization matrix}
			A_j & = \left.\frac{\partial f_j}{\partial x}\right\vert_{\substack{x=0\\w=0}}, & B_j^{(1)} & = \left.\frac{\partial f_j}{\partial u}\right\vert_{\substack{x=0\\w=0}},  & B_j^{(2)} & = \left.\frac{\partial f_j}{\partial w}\right\vert_{\substack{x=0\\w=0}}  \nonumber \\
			C_j & = \left.\frac{\partial g_j}{\partial x}\right\vert_{\substack{x=0\\w=0}}, & D_j & = \left.\frac{\partial g_j}{\partial w}\right\vert_{\substack{x=0\\w=0}}, & &
		\end{align}
		\vspace{-3pt}
\begin{equation}
{H = \begin{bmatrix}
\frac{\partial u_1}{\partial x_1} & \cdots & \frac{\partial u_1}{\partial x_{N}}\\
\vdots & \vdots &\vdots\\
\frac{\partial u_{N}}{\partial x_1} & \cdots & \frac{\partial u_{N}}{\partial x_{N}}
\end{bmatrix}}_{x=0,w=0} \text{where}\end{equation}
{$$
{\frac{\partial u_i}{\partial x_k} = \begin{bmatrix}
\frac{\partial \Delta P^i_{inj}}{\partial \Delta \delta_k} & \frac{\partial \Delta P^i_{inj}}{\partial \Delta \omega_k} & \frac{\partial \Delta P^i_{inj}}{\partial \Delta V_k} \\[5pt] 
\frac{\partial \Delta Q^i_{inj}}{\partial \Delta \delta_k} & \frac{\partial \Delta Q^i_{inj}}{\partial \Delta \omega_k} & \frac{\partial \Delta Q^i_{inj}}{\partial \Delta V_k}
\end{bmatrix}}, {i,k \in \{1,.., N\}}.$$} 
\end{subequations}
 The matrix $H$ is the power flow Jacobian corresponding to the linearization of \eqref{powerflow}. 

Equation \eqref{nonlinear_switched} represents the dynamics of the network of interconnected microgrids with MAFD primary control, and \eqref{linear switched system} is its linear approximation. The deployment of MAFD control to address intermittent loss of angle measurement inherently induces switched dynamics in the system. This calls for a secondary controller which should not only handle the deviations caused by generation-load mismatches in the microgrid but also stabilize the network of microgrids during the switching transients introduced by the MAFD controller. In this paper, we design a switched secondary controller to address the problem of transient stability. This problem is stated more formally as follows.
\vspace{5pt}
\begin{problemstatement}[Transient Stability]
Given the linearized switched system model \eqref{linear switched system}, design a \textit{secondary} output-feedback control input $\tilde u(t)=K_{\sigma(t)}y(t)$, ${K_j \in \mathbb{R}^{2N\times2N}}$, $j \in \Sigma^N$, such that the nonlinear switched system \eqref{nonlinear_switched} with $u(t) \mapsto u(t)+\tilde{u}(t)$ is locally stable with respect to $w(t)$ (in the sense of $\mathcal{L}_2$ stability) for any switching between angle (and voltage) droop and frequency (and voltage) droop primary controllers of individual microgrids in the network. 
\end{problemstatement}

\section{Secondary Control Synthesis}
\label{sec:solution}
We now present a secondary control design based on the notion of $QSR$-dissipativity \cite{hill1980dissipative} 
 for the MAFD framework discussed in Section \ref{sec:mafd-model}, with the aim of ensuring transient stability when the system switches between angle droop and frequency droop control modes. The proofs of the results presented in this section are omitted due to space constraints.
%
We begin by presenting some  definitions and results that will be used in this work.

\begin{definition}\label{def:QSR}
A switched system \eqref{nonlinear_switched} is said to be $QSR$-dissipative with input $w$ and dissipativity matrices $Q_j$, $S_j$ and $R_j$, $j \in \Sigma^N$, if there exists a positive definite storage function $V(x):\mathbb{R}^{3N}\rightarrow\mathbb{R}_+$ such that for all $t\geq t_0 \geq 0$,
\begin{equation*}\label{QSR1}
\int_{t_0}^{t} \begin{bmatrix}
y(\tau) \\
w(\tau)
\end{bmatrix}'\begin{bmatrix}
Q_j & S_j\\
S_j' & R_j
\end{bmatrix}\begin{bmatrix}
y(\tau) \\
w(\tau)
\end{bmatrix} d\tau \geq V(x(t))-V(x(t_0))
\end{equation*}
holds, where $x(t)$ is the state at time $t$ resulting from the initial condition $x(t_0)$ and input $w(\cdot)$. 
Additionally, \eqref{nonlinear_switched} is said to be \textit{$QSR$-state strictly input dissipative} ($QSR$-SSID) if, for all $t\in \mathbb{R}_+$ and $j \in \Sigma^N$,  
{\small
\begin{align*}\label{QSR-SSID}
\begin{bmatrix}
y(t) \\
w(t)
\end{bmatrix}'\begin{bmatrix}
Q_j & S_j\\
S_j' & R_j
\end{bmatrix}\begin{bmatrix}
y(t) \\
w(t)
\end{bmatrix} &\geq \dot{V}(x(t))+ \phi_j(w(t)) + \psi_j(x(t)),
\end{align*}}where $\phi_j(\cdot),\psi_j(\cdot)$ are positive definite functions of $w(t)$ and $x(t)$ respectively. 
A switched system \eqref{nonlinear_switched} is said to be locally $QSR$-dissipative if it is $QSR$-dissipative for all $x\in\mathcal{X}$ and $w\in\mathcal{W}$ where $\mathcal{X}\times\mathcal{W}$ is a neighborhood of $x,w=0$.
\end{definition}

$QSR$-dissipativity is closely related to input-output stability of the switched system and can also be used to capture several other properties such as robustness and transient performance via appropriate choice of the $Q_j$, $S_j$ and $R_j$ matrices \cite{agarwal2019compositional}. 
A $QSR$-dissipative switched system \eqref{nonlinear_switched} is $\mathcal{L}_2$ stable if $Q_j<0$ for every $j \in \Sigma^N$.

\subsection{Design Equations}\label{sec:design}
The local stability of closed loop system can be guaranteed by choosing the control input $\tilde{u}(t)$ such that \eqref{nonlinear_switched} is locally $QSR$-dissipative for $u(t) \mapsto u(t)+\tilde{u}(t)$, with $Q_j<0$, $j \in \Sigma^N$. However, it can be quite difficult to design such a control input for nonlinear systems, especially with the added complexity resulting from switching dynamics. {Therefore, we develop a control design to ensure local $QSR$-dissipativity of the nonlinear switched system \eqref{nonlinear_switched} from that of its first order linear approximation \eqref{linear switched system}.}
\begin{theorem}\label{thm:design}If for all $j \in \Sigma^N$, $B_j^{(1)}$ is full column rank and there exists symmetric positive definite matrix \(P \in \mathbb{R}^{ 3N\times 3N}\) and matrices \(U_j, V_j\) of appropriate dimensions such that \eqref{control_lmi} holds, where $\mathcal{S}_v$ is the set of all diagonal matrices and $\mathcal{S}_H$ is the set of all matrices with the same sparsity structure as the Jacobian matrix $H$ in \eqref{linear switched system}, 
then the output feedback control law $u(t) \mapsto u(t) + \tilde{u}(t)$ where \(\tilde{u}(t)=K_{\sigma(t)}y(t)\) with $K_j=V_j^{-1}U_j, \forall j \in \Sigma^N$ renders the system \eqref{nonlinear_switched} locally $QSR$-dissipative, and hence locally $\mathcal{L}_2$ stable, for any switching sequence. The dissipativity matrices for closed loop system are given by $Q_j<0$, $S_j$ and $R_j$, $j \in \Sigma^N$.
\end{theorem}

Theorem \ref{thm:design} provides control design equations to ensure local stability of the nonlinear switched system \eqref{nonlinear_switched}, based on its linearized model \eqref{linear switched system}. Note that the assumption that $B_j^{(1)}$ has full column rank implies that all inputs affect the output in a linearly independent manner, that is, there are no redundant control inputs. This assumption is sufficiently general since redundant control inputs, if present, can be combined to achieve full column rank.


\begin{remark} We make the following comments about the proposed control synthesis.
\begin{itemize}
\item[(i)] The constraints on the sparsity structure of matrices $V_j$ and $U_j$ in \eqref{pass_lmi2} guarantee that the secondary controller gain matrix designed using the results in Theorem \ref{thm:design} is distributed, wherein each microgrid only uses output measurements from its immediate neighbors, thereby reducing the communication overhead.
\item[(i)] The design equations in \eqref{control_lmi} are provided in the form of LMIs, rather than the nonlinear matrix inequalities typically encountered in dissipativity-based designs for nonlinear switched systems \cite{li2016dissipativity}\cite{pang2016incremental}. 
\item[(ii)] Note that the results in Section \ref{sec:design} are more generally applicable to any nonlinear switched system, 
and not restricted to interconnected microgrids of the form \eqref{nonlinear_switched}.
\end{itemize}
\end{remark}

\section{Case Study}
\begin{figure}[b]
	\vspace{-1em}
			\centering
			\includegraphics[scale=0.3,trim=1.1cm 0.7cm 0.6cm 0cm]{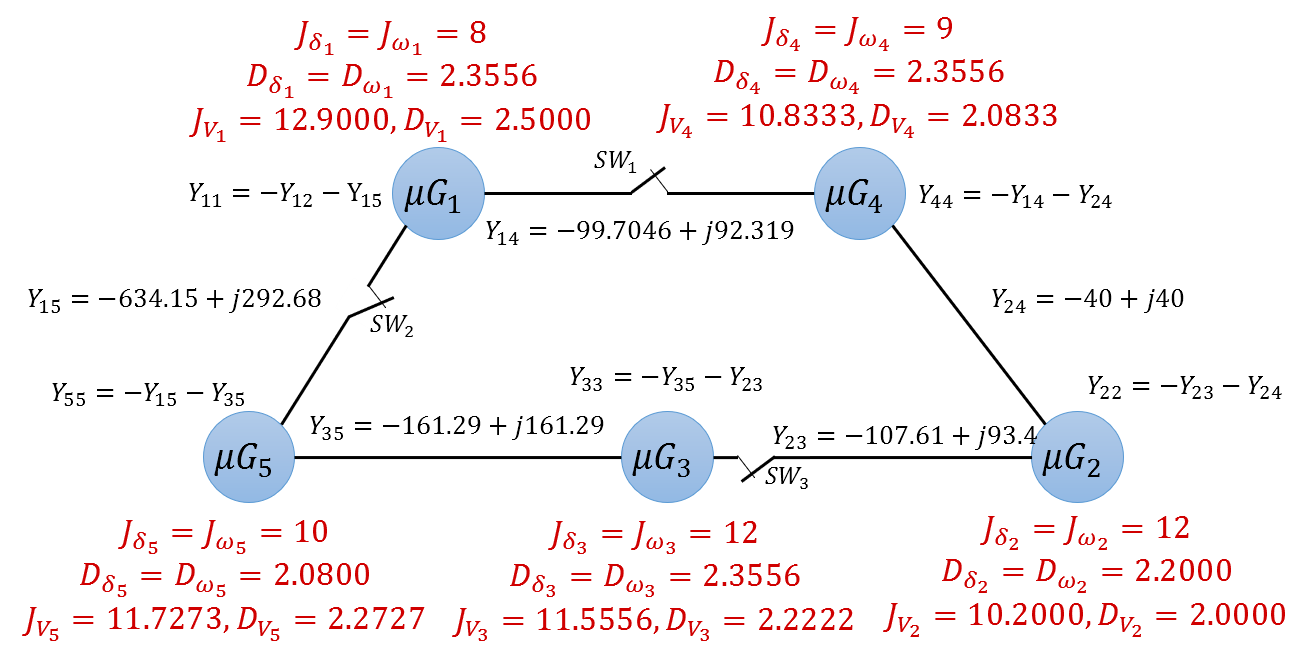}
			\caption{Network parameters (p.u.) for 123-feeder five-microgrid test system.}
			\label{fig:network_param}
	\end{figure}
	
	\begin{figure*}[t]
		\centering
		   \vspace{0.5em}
		\includegraphics[scale=0.6,trim=0.1cm 0.4cm 0cm 0.2cm]{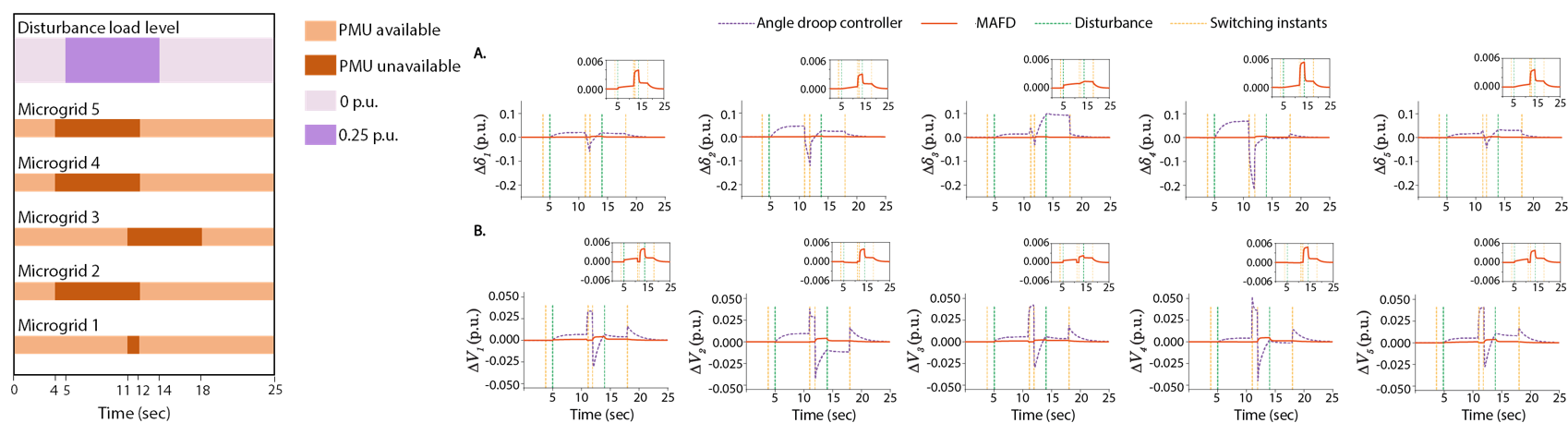}
		\caption{Left: Disturbance  $w(t)$ and switching signal $\sigma(t)$ corresponding to PMU angle measurement loss for five-microgrid test system. Right: A. Angle errors, and B. voltage errors of the MAFD design (\textbf{C1}) compared with a traditional angle droop controller (\textbf{C2}) for the disturbance $w(t)$ (Left).}
		\label{fig:case}
	\end{figure*}
In this case study, we consider a five-microgrid system, 
as shown in Fig. \ref{fig:network_param}. To illustrate the performance of the MAFD framework, we consider the two following set-ups:
\begin{itemize}
    \item \textbf{C1 (MAFD framework):} The primary controller here is the proposed mixed voltage angle and frequency droop control. We obtain a nonlinear switched system model of the form \eqref{nonlinear_switched} with 32 switching modes, i.e., $\sigma(t) \in \Sigma: \{1,2\}\times\{1,2\}\times\{1,2\}\times\{1,2\}\times\{1,2\}$, and linearize it around the power flow operating point in Table \ref{operating_point_five}. We use this linearized model to design a stabilizinhg distributed output-feedback secondary controller obtained by solving \eqref{control_lmi}, and $K_j=V_j^{-1}U_j, \forall j \in \Sigma^N$.
    \item \textbf{C2 (Angle droop control):} In this setup, we assume that all microgrids use angle droop control as the primary controller and continue to use the same with the last available measurement even when angle measurements are lost, that is, with the dynamics corresponding to the mode $j=[1 \, 1 \ldots \, 1]$ in \eqref{nonlinear_switched}. We design a centralized secondary controller obtained by solving \eqref{pass_lmi1}, \eqref{pass_lmi2} and $K_j=V_j^{-1}U_j, j=[1 \, 1 \ldots \, 1]$. 
\end{itemize}

We compare the performance of the two controllers \textbf{C1} and \textbf{C2} by simulation on the original nonlinear system \eqref{nonlinear_switched} for a test pattern of angle measurement losses and disturbance as shown in Fig. \ref{fig:case}-Left acting on all microgrids. 
From the angle and voltage profiles (Fig. \ref{fig:case}-Right), we observe:
\begin{itemize}
    \item The MAFD controller successfully stabilizes the system under the measurement loss and disturbance pattern with significantly improved transient performance as compared to just the angle droop controller with secondary controller using the last available angle measurement in the event of measurement loss. 
    \item The voltage profiles resulting from both controllers are similar, since voltage magnitude measurements continue to be available, and the voltage droop control loop is largely unaffected.
\end{itemize}
	\vspace{-0.4em}
\section{Conclusion}
\label{sec:conclusions}
We presented a mixed angle-frequency droop control (MAFD) framework for interconnected microgrids where angle measurements may be intermittently lost, and proposed a dissipativity-based secondary control design that guarantees transient stability. Besides scenarios of PMU measurement loss, the proposed MAFD framework is also more generally applicable in legacy systems where some microgrids operate with angle droop control and others continue to use traditional frequency droop control.
\begin{table}[t]
\centering
\vspace{-1em}
\caption{Power flow solution for 123-bus 5-microgrid test system}
\label{operating_point_five}
\begin{tabular}{lllllll}
\hline
          & $P_{inj}^{ref}$ & $Q_{inj}^{ref}$ & $P_{load}^{ref}$ & $Q_{load}^{ref}$ & $V^{ref}$ & $\delta^{ref}$    \\
        & (p.u.) & (p.u.) & (p.u.) & (p.u.) & (p.u.) & (deg.)
          \\
          \hline
$\mu G_1$ & 0.79   &   1.35   &   0.92   &   0.47   & 1.000     & 0.000 \\
$\mu G_2$ & 0.80   &   0.10   &   0.23   &  0.11    & 1.003  &  0.233 \\
$\mu G_3$ & 0.20   &  0.10    &  0.45    &  0.20     & 1.000 & 0.110 \\
$\mu G_4$ & 0.80   &   0.20   &  0.27    &  0.12    & 1.003  &  0.158  \\
$\mu G_5$ & 0.20   &   0.10   &   0.92   &   0.95    & 0.999 &  0.052 \\\hline
\end{tabular}
\vspace{-1em}
\end{table}
\vspace{-0.0em}

\bibliographystyle{IEEEtran}
\bibliography{references}  

\end{document}